\documentclass[10pt,a4paper,leqno]{article}
\usepackage[english]{babel}
\usepackage[latin1]{inputenc}
\usepackage{graphicx}
\usepackage{epsfig}
\usepackage{amsmath, amssymb}
\usepackage{color}
\usepackage{stmaryrd}
\usepackage{fancyhdr}
\usepackage{vmargin}
\setmarginsrb{1.8cm}{2cm}{1.8cm}{2cm}{1cm}{1cm}{1cm}{1cm}
\numberwithin{equation}{section}

\definecolor{rosso}{rgb}{1,0,0}
\definecolor{nero}{cmyk}{0,0,0,1}
\definecolor{blu}{rgb}{0,0,1}

\newtheorem{remarque}{Remark}[section]
\newtheorem{definition}{Definition}[section]

\newtheorem{exemple}{Example}[section]

\title{Risk Premium Impact in the Perturbative Black Scholes Model}

\author{\linespread{1} Luca Regis\footnote{University of Torino  \newline
  address: Via Real Collegio 30, 10024 Moncalieri (TO) Italy \newline email:
  luca.regis@unito.it} $\; \;$ and Simone Scotti\footnote{University of Torino and Ecole Nationale
  des Ponts et Chaussées - CERMICS    \newline
  address: Via Real Collegio 30, 10024 Moncalieri (TO) Italy  \newline email:
  simone.scotti@unito.it}}
\date{ }

\begin{document}

\maketitle
\color{nero}

\begin{abstract}
  
We study the risk premium impact in the Perturbative Black  Scholes model.
The Perturbative Black Scholes model, developed by  Scotti, is a subjective
volatility model based on the classical  Black Scholes one, where the
volatility used by the trader is  an estimation of the market one and
contains measurement errors.  In this article we analyze the correction to
the pricing formulas  due to the presence of an underlying drift different
from the risk free return.  We prove that, under some hypothesis on the
parameters, if  the asset price is a sub-martingale under historical
probability, then  the implied volatility presents a skewed structure, and
the  position of the minimum depends on the risk premium $\lambda$.

%\subclass{60H30 \and 91B16 \and 91B70}


\end{abstract}

\section{Introduction}

It is now common knowledge that the Black and Scholes model, which worked
well before the 1987 crash, is nowadays unable to price options correctly.
As can be deduced by comparing the two papers by Rubinstein \cite%
{bib:Rubinstein} and Jackwerth and Rubinstein \cite{bib:Jackwerth_Rubinstein}%
, something has evidently changed in the market for options after that
event. The most shared explanation for the failure in which BS model incurs
today is usually thought to reside in the fact that the constant volatility
parameter it proposes is not a good representation of reality anymore.
Empirical evidence shows that the underlying stock volatility, for example,
is not time invariant during the life of an option.

Moreover, while before 1987 the lognormal distribution of stock prices
implied by the Black and Scholes model seemed to be a good approximation of
the real one and volatility observed across strike prices had a moderately
pronounced smile, from that date onwards the implied volatility curve
appears to be steeper and generally skewed to the left. Jackwerth and
Rubinstein \cite{bib:Jackwerth_Rubinstein}, recovering stock price
distribution from observed prices, empirically find a ``fatter'' left tail
phenomenon. Constantinides et al. \cite{bib:Constantinides}, in the context
of an equilibrium model, find stochastic dominance violations on both tails
of the implied volatility curve. Christensens and Prabahla \cite%
{bib:Christensen}, for example, suggest that a regime switch has occurred
after the crash.

The most natural solution to the problem of pricing options more correctly
seems then to let volatility change with time. Many pricing models have then
been proposed, with different formulations for the stochastic process
driving volatility, e.g. local volatility models, see Dupire \cite%
{bib:Dupire}, or stochastic volatility models. Then, an option pricing model
is characterized by a system of differential equations, since two different
processes are specified, one for the stock price and one for its underlying
volatility.

The first authors to solve the problem of pricing options with stochastic
volatility were Hull and White \cite{bib:Hull_White}. However, they were
able to obtain closed form solutions only for the case of uncorrelated
volatilities and stock prices, while Heston \cite{bib:Heston}, using a new
technique, managed to find exact prices also for the correlated case. In
this stream of literature, one of the most used model in practice is
probably the SABR one, introduced in Hagan et al. \cite{bib:SABR}, which
provides excellent fitting for interest rates derivatives. More
sophisticated models include for example the possibility of jumps in the
stock price evolution, e.g. see Brigo et al. \cite{bib:Brigo-Mercurio}, or
directly in the process for volatility, see Eraker et al. \cite{bib:Eraker}.

However, stochastic volatility models like the ones described above are
usually complex and characterized by a large number of parameters and,
unless in special cases (SABR\ model, Heston \cite{bib:Heston}), they do not
provide closed form solutions for vanilla options prices.

The PBS model (Scotti \cite{bib:Scotti-PBS}) introduced a new category of
stochastic volatility pricing models, being founded on the notion of
subjective volatility. Using error theory through Dirichlet forms, the PBS
model generalizes the standard Black and Scholes one, imposing an error
structure on volatility. Thus, rather than specifying a possible pattern of
evolution for volatility through time, the perturbative approach deals with
the concept of measurement errors present in the estimation procedure
performed by the trader.

One of the most important issues of this model lies in the possibility of
obtaining closed forms solutions for European vanilla option prices and for
each kind of derivative which has a closed form solution using the classical
Black and Scholes model such as Asian and barrier options. This important
framework, joint with the flexibility of the model, permits us to calibrate
it to different markets and fit them, even if they imply opposite behaviours
of the implied volatility curve. PBS\ can reproduce a right-tailed or a
left-tailed skewness effect, as well as sharper or flatter slopes, obviously
depending on the calibration of parameters. In Scotti \cite{bib:Scotti-PBS},
sufficient conditions for the presence of a smile are derived in the case
with no drift term in the stock price dynamics.

This article studies a natural implication of the PBS\ model: the dependence
of option prices on risk premia. Dependence of option prices on the expected
excess return on the stock is ruled out in the classical Black and Scholes
model, as a result of the lognormality assumption. Lo and Wang \cite%
{bib:Lo_Wang}, starting from the evidence that the predictability naturally
captured by the expected return on stocks is affects option prices, assume
an Orstein-Uhlenback process for stock prices and show that the impact of a
drift term is not negligible anymore.

In the PBS model we maintain the assumption of lognormality of stock prices.
However, we show that the presence of measurement errors in the estimation
of parameters induces market incompleteness and lets the hedging position of
a trader not invariant to the expected excess return on stock. We analyze
the implications of this fact and we show that taking into account the
impact of a risk premium on stocks it is possible to reproduce the most
commonly observed behaviours of the market for options in terms of implied
volatility. The PBS\ model is able to generate the usual smile and skew
effects pointed out by the empirical literature we addressed above.

Section 2 briefly summarizes the basic concepts of error theory based on
Dirichlet forms, section 3 recalls the most important features of the PBS\
model, section 4 studies the impact of the drift term on the general profit
and loss function and in the particular case of the price of a European call
option, section 5 analyzes the sensitivity of the volatility implied by the
model to some parameters and above all to risk premium. Finally, the
appendix provides technical computations.

\bigskip

\section{Preliminaries}

This section resumes the notations used throughout the paper and briefly
surveys the key notions of error theory using Dirichlet forms.\footnote{%
For a complete and  exhaustive treatment of error theory and Dirichlet
forms, see Bouleau  \cite{bib:Bouleau-Hirsch}, \cite{bib:Bouleau-erreur2}, 
\cite{bib:Bouleau-erreur4}  and Fukushima \cite{bib:Fukushima}.}

\subsection{Notation}

We use the following notation:

\begin{itemize}
\item $\left(\Omega, \, \mathcal{F}, \, \mathbb{P}\right)$ is the historical
probability space, sometimes simply denoted with $\Omega$.

\item $\left\{\mathcal{F}_t\right\}_{0 \leq t \leq T}$ is a filtration
defined on the  probability space.

\item $\left\{W_t\right\}_{0 \leq t \leq T}$ is the associated Brownian
motion, i.e.  a Brownian motion adapted to the filtration $\left\{\mathcal{F}%
_t\right\}_{0 \leq t \leq T}$.

\item $\left(\widetilde{\Omega}, \, \widetilde{\mathcal{F}}, \,  \widetilde{%
\mathbb{P}}\right)$ is another probability space, used to represent  the
uncertainty on the volatility parameter, denoted simply with  $\widetilde{%
\Omega}$.  %\item $\left(\widehat{\Omega}, \, \widehat{\mathcal{F}}, \,
%\widehat{\mathbb{P}}\right)$ is a copy of the probability space
%$\widetilde{\Omega}$, i.e. a probability space independent of
%$\widetilde{\Omega}$, but with the same sigma-algebras and %probability laws.

\item $\mathbb{E}[\bullet]$ and $\mathbb{E}[\bullet|\mathcal{F}_t ]$ denote,
respectively, the expectation and the conditional expectation under the
probability  measure $\mathbb{P}$, while $\widetilde{\mathbb{E}}[\bullet]$
denotes the  expectation under $\widetilde{\mathbb{P}}$.  
%, while %$\widehat{\mathbb{E}}[\;\;\;]$
%denotes the expectation under the %probability $\widehat{\mathbb{P}}$

\item $\left(P_t\right)_{t \geq 0}$ denotes a strongly continuous
contraction  semi-group, $\mathcal{A}$ its generator, with domain $\mathcal{%
DA}$, and $\Gamma$  the "carré du champ" operator associated with the
Dirichlet form of the semi-group,  with domain $\mathbb{D}$.

\item $T$ denotes the maturity, a fixed positive number.

\item we suppose that there are two traded assets in the market: a stock,
with price  $S_t$ at time t, and a risk free money account. For the sake of
simplicity we  assume that the risk free interest rate is zero.

\item $\sigma_0$ is the stock market volatility, supposed to be constant,
while  $\sigma$ is the volatility estimated by the trader which we assume to
be  stochastic.

\item $\mu$ is the the expected rate of return on the stock, $\lambda$
represents the  risk premium and $\mathcal{L}$ the cumulated risk premium,
i.e. $\mathcal{L}=  \lambda \, \sqrt{T}$
\end{itemize}

\subsection{Error Theory using Dirichlet Forms}

Asset pricing models are characterized by a set of parameters that permits
to reproduce the real world features with some degree of freedom. These
parameters have to be estimated in order to be used in practice. The
classical way to find these estimations is to use market data to make the
model fit the prices observed on the market. Hence, the final result of this
procedure is that we obtain a mean value for each parameter along with some
uncertainty due to estimation errors.

What is the impact of such an uncertainty? Generally, models' users consider
the mean values of the estimations of their parameters and implicitly
consider them as deterministic, since their variance is generally small with
respect to their mean. However, when non linear functions are then used in
pricing procedures, the impact of this uncertainty, even if small, could be
important, since the moments of random variables are distorted. What is the
amplitude of the bias induced by forgetting the probabilistic nature of an
estimated parameter?

Giving an answer to this question is not an easy task. If we consider a non
linear function $F$ and a random variable $\sigma$ it is well know by
Jensen's inequality that the expected value of $F(\sigma)$ is different from
the function F evaluated at the expected value of $\sigma$. A
straightforward computation of this difference, however, is usually
impossible, since the functional forms involved are often difficult to
treat. This is the reason why the probabilistic uncertainty on the
estimation of parameters, whose presence was already known by Gauss, is
frequently neglected.

If a direct approach to solve this problem is unsuccessful, we can try to
find a way to reach a solution addressing the question from another
perspective. We can use the fact that the variance of an estimated parameter
is very small compared to its mean in order to justify a Taylor expansion;
studying the bias and the variance of $F(\sigma)$ we find:

\begin{equation}  \label{eqn:chain-rule}
\begin{array}{rcl}
\widetilde{ \mathbb{E}}\left[F(\sigma) - F(\sigma_0) \right] & = & 
\epsilon\left\{ F^{\prime }(\sigma_0) \; \text{ Bias }[\sigma] +\frac{1}{2}
\, F^{\prime \prime }(\sigma_0)\; \text{ Variance }[\sigma] \right\} +
o(\epsilon) \\ 
\widetilde{\mathbb{E}}\left[\left(F(\sigma) - F(\sigma_0) \right)^2 \right]
& = & \epsilon \, \left(F^{\prime }(\sigma_0)\right)^2\; \text{ Variance }%
[\sigma] + o(\epsilon)%
\end{array}%
\end{equation}
where $\sigma_0$ is the estimated value of the random variable $\sigma$, $%
\epsilon \text{ Bias }[\sigma]$ is its bias, i.e. the difference between $%
\sigma_0$ and $\widetilde{\mathbb{E}}[\sigma]$, and $\epsilon \text{
Variance }[\sigma]$ is its variance.

\begin{remarque}
The first relation in (\ref{eqn:chain-rule}) summarizes the key  difference
between a deterministic uncertainty and a  probabilistic one, since the mean
of a function  evaluated on a random variable is distorted if it is non
linear.
\end{remarque}

The starting point of error theory using Dirichlet forms consists in
considering a very small $\epsilon$ and to stop the Taylor expansion at the
first order. The theory involves thus the use of two operators, a bias
operator $\mathcal{A}$ and a variance-covariance one $\Gamma$, linked
together by the chain rule in equation (\ref{eqn:chain-rule}). Such an
environment can be studied through semi groups theory. The operator $%
\mathcal{A}$ is the generator of the semi-group and $\Gamma$ is the "carré
du champ" associated with the Dirichlet form of the semi-group.

The most important framework of error theory is the error structure. We
recall its definition:

\begin{definition}[Error structure]

  An error structure is a term

  \begin{displaymath}
  \left( \widetilde{\Omega}, \, \widetilde{\mathcal{F}},
  \, \widetilde{\mathbb{P}}, \, \mathbb{D}, \, \Gamma \right)
\end{displaymath}

where

\begin{enumerate}

\item{$\left( \widetilde{\Omega}, \, \widetilde{\mathcal{F}}, \,
      \widetilde{\mathbb{P}} \right)$ is a probability space;}

\item{$\mathbb{D}$ is a dense sub-vector space of $L^2\left( \widetilde{\Omega}, \,
      \widetilde{\mathcal{F}}, \, \widetilde{\mathbb{P}}\right)$;}
\item{$\Gamma$ is a positive symmetric bilinear application from $\mathbb{D} \,
    \times \, \mathbb{D}$ into $L^1 \left( \widetilde{\Omega}, \,
      \widetilde{\mathcal{F}}, \, \widetilde{\mathbb{P}} \right)$ satisfying the
    functional calculus requirements on the class $\mathcal{C}^1 \cap Lip$. This
    means that, if F and G are two functions of class $\mathcal{C}^1$ and
    Lipschitzian, u and v $\in \mathbb{D}$, then F(u) and G(v) $\in \mathbb{D}$ and}

  \begin{displaymath}
  \Gamma\left[F(u), \, G(v) \right] = F'(u) G'(v) \Gamma[u, \, v] \; \;
  \widetilde{\mathbb{P}} \; a.s.;
\end{displaymath}

\item{the bilinear form $\displaystyle \mathcal{E}[u, \, v] = \frac{1}{2}
    \widetilde{\mathbb{E}}\left[\Gamma[u, \, v]\right]$ is closed;}

\item{The constant function 1 belongs to $\mathbb{D}$, i.e. the error structure is
    Markovian.}

\end{enumerate}

\end{definition}
Hypotheses 2, 3 and 4 together ensure that $\mathcal{E}$ is a Dirichlet
form, with $\Gamma$ as carré du champ operator.

We use the simplified notation $\Gamma[u] = \Gamma[u, \, u]$ to indicate
that the operator $\Gamma$ is applied twice on the same argument. The couple 
$(\Gamma, \widetilde{\mathbb{P}})$ defines a unique semi group $%
\displaystyle
\left(P_t\right)_{t \geq 0}$ and its generator $\mathcal{A}$ thanks to the
Hille-Yosida theorem \footnote{%
see Fukushima \cite{bib:Fukushima} for a complete  proof}. Hypothesis 5
provides that the semi-group $\left(P_t\right)_{t \geq 0}$ is Markovian.

Therefore, we defined two operators $\Gamma$ and $\mathcal{A}$ that satisfy
the chain rule (\ref{eqn:chain-rule}). \newline
It is now useful to conclude this section providing an example of an error
structure:

\begin{exemple}[Orstein-Uhlenbeck structure]
\begin{equation*}
\left(\widetilde{\Omega}, \, \widetilde{\mathcal{F}}, \, \widetilde{\mathbb{P%
}}, \,  \mathbb{D}, \, \Gamma \right) = \left(\mathbb{R}, \, \mathcal{B}(%
\mathbb{R}), \,  \mu, H^1(\mu), \Gamma[u, \,u] = \left\{u^{\prime}\right\}^2
\right) 
\end{equation*}
where $\mathcal{B}(\mathbb{R})$ is the Borel $\sigma$-field of $\mathbb{R}$, 
$\mu$ is a gaussian measure and $H^1(\mu)$ is the first Sobolev space with
respect to the measure $\mu$, i.e. $u \in H^1(\mu)$ if $u \in L^2(\mu)$ and $%
u^{\prime}$ belongs to $L^2(\mu)$ in distribution sense.

The associated generator has the following domain:

\begin{equation*}
\mathcal{D}\mathcal{A} = \left\{u \in L^2(\mu): \, u^{\prime \prime} - x
\,f^{\prime}  \text{ belongs to } L^2(\mu)\right\} \text{in distribution
sense} 
\end{equation*}

and the generator operator is

\begin{equation*}
\mathcal{A}[u] = \frac{1}{2} u^{\prime \prime} - \frac{1}{2} I \cdot
u^{\prime} 
\end{equation*}

where $I$ is the identity map on $\mathbb{R}$.
\end{exemple}

This example gives the basic idea of an error structure on a parameter.
Moreover, as shown in Hamza \cite{bib:Hamza}, it is important since every
Dirichlet form on $\mathbb{R}$ has a characterization.

\section{Perturbative Black Scholes model}

In this section, we recall the key features of the Perturbative Black
Scholes model introduced by Scotti \cite{bib:Scotti-PBS}.

The PBS model is based on the classical Black Scholes model, (Black and
Scholes \cite{bib:Black-Scholes}). If we assume that the interest rate is
worth zero or, from an economic point of view, that all assets are priced in
terms of the money market, then the underlying stock price follows the SDE

\begin{equation}  \label{eqn:SDE-BS}
dS_t = \mu \, S_t \, dt + \sigma_0 \;S_t \; dW_t
\end{equation}
where $\mu$ is the return on the stock, $\sigma_0$ is the volatility and $W_t
$ is a Brownian motion.

In the BS model pricing formulas depend on the diffusion term only and not
on $\mu$; we find closed forms expressions for the premium and the greeks of
vanilla options. In contrast with its simplicity, unluckily the BS model
cannot reproduce the so called smile effect: the volatility implied by the
BS model is constant across strike prices, while the observed one is usually
u-shaped.

We introduce the notation of risk premium $\lambda$ as the ratio between the
expected return on the stock and market volatility.

\begin{equation*}
\lambda = \frac{\mu}{\sigma_0} 
\end{equation*}

The PBS model lies on three main hypotheses:

\begin{enumerate}
\item {\ the stock price follows a geometrical brownian motion with fixed
and non  perturbed volatility $\sigma_0$};

\item {\ the trader has to estimate the volatility parameter, and the value
of his own  estimation contains intrinsic inaccuracies. The model reproduces
this fact  through an error structure; nonetheless we assume that the stock
price $S_t$ is  not erroneous. We evaluate the impact of the perturbation
generated by those  measurement errors on the profit and loss process used
by trader to hedge a  position on a vanilla option};

\item {\ the trader knows the existence of the perturbation described above
and wants  to modify his own offered prices in order to take into account
the bias present  on volatility and, as a consequence, on the hedged position%
}.
\end{enumerate}

Summarizing, all traders use a geometric Brownian motion to model the stock
price process and they hold some positions involving vanilla options; they
use observed market prices to determine the values of parameters by
inversion of pricing formulas. Thus, they find an observed volatility
process $\varsigma_t$, usually known as implied volatility, they take it as
a forecast for future volatility and hedge their portfolio accordingly.

Since we made as an assumption that the trader knows the existence of errors
in his estimation procedure, volatility is incorporated into the model in
two different ways. It has a ``market'' value, the classical parameter used
to set up standard pricing formulas, and a subjective one. The former is
denoted with $\sigma_0$ and it is supposed to be a constant parameter as in
the Black Scholes model. The ``subjective'' volatility notion comes from the
intuition that in the real world, when an operator deals with the problem of
option pricing, he does not know the precise value volatility will assume
during its life. Hence, he has to estimate it from market observations. The
value he gets from this procedure, as pointed out above, will obviously be
subject to measurement errors\footnote{
Measurement errors arise from the uncertainty expected using the central
limit  theorem.}, captured in the PBS\ model by the error structure form. We
assume that the stock volatility ``market'' value is also the mean value of
volatility in the erroneous estimation procedure performed by the trader.
The volatility estimated by the trader is then a random variable, and is
"subjective", since it can assume a different value in the expectation of
each operator.

The profit and loss process of a trader has a key role in the PBS model. The
value of this process at maturity is given by:

\begin{equation}  \label{equation:P-and-L}
P\&L = F(\varsigma_0, \, S_0, \, 0) + \int_0^T\frac{\partial F}{\partial x}%
(\varsigma_t, \, S_t, \, t) dS_t - \Phi(S_T)
\end{equation}

where $F(\varsigma_0, \, S_0, \, 0)$ is the security premium, the integral
term represents the hedging strategy, $\Phi(S_T)$ is the Payoff and $S_t$
follows Black Scholes SDE (\ref{eqn:SDE-BS}).

\section{Impact of the drift term in security pricing}

In this section, we study the impact of a non zero drift term in the
diffusive process assumed for stock prices on prices determined with the PBS
model.

As we have shown previously, the expected profit and loss function from the
hedging position is then in turn a random variable, characterized by a bias
and a variance term, which make it different from the one implied by the BS\
model. We make an important remark:

\begin{remarque}[Drift impact]
In the PBS model, the profit and loss process defined in equation (\ref%
{equation:P-and-L}) depends crucially on the drift rate $\mu$, which is the
expected excess return on the stock. As a matter of fact, the integral term
in (\ref{equation:P-and-L}) depends on the diffusive process described in (%
\ref{eqn:SDE-BS}) where $\mu$ plays a role.

In the Black Scholes model, instead, the price of an option does not depend
on the drift term. In that case, in fact, the $P \& L$ process is worth zero
almost surely; as a consequence, we can change the probability measure
without altering the result. If, as in the PBS model, we assume that the
volatility $\varsigma_t$ used by the trader is $\sigma$ and not $\sigma_0$,
the profit and loss process is not worth zero a.s.; on the contrary, it
becomes a stochastic process characterized by two random sources:

\begin{itemize}
\item the Brownian motion which describes the evolution of the stock price
and

\item the process $\varsigma_t$, the trader's volatility, which depends on
an  independent probability space.
\end{itemize}

As a consequence, we cannot change the probability measure without changing
the value of the profit and loss process at maturity.
\end{remarque}

We suppose that the trader's volatility $\varsigma_t$ is the time
independent random variable $\sigma$ we defined in the previous section.
Using the language of Dirichlet forms, we derive the following expansion for
the volatility estimation:

\begin{equation*}
\sigma_0 \rightarrow \sigma_0 + \epsilon \mathcal{A}[\sigma](\sigma_0) + 
\sqrt{\epsilon \Gamma[\sigma](\sigma_0)} \widetilde{\mathcal{N}} 
\end{equation*}
where $\widetilde{\mathcal{N}}$ is a standard Gaussian random variable.
Moreover, we assume that this error structure admits a sharp operator.

We estimate the variance and bias of the error on $\mathbb{E}\left[P \& L%
\right]$. In the computation we assume that $\sigma = \sigma_0$ is the right
value of the random variable, in the sense that if $\varsigma_t=\sigma_0$,
then $P \& L(\sigma_0) =0$ almost surely. Notice that, however, this does
not mean that the trader believes the BS model to be correct.

Then we can prove, see Scotti \cite{bib:Scotti-PBS}, that we have the
following bias and variance terms:

\begin{equation}  \label{relation-bias-variance}
\begin{array}{rcl}
\displaystyle \mathcal{A} [\mathbb{E}[P \& L]] & = & \displaystyle \left\{%
\frac{\partial F}{\partial \sigma}(\sigma_0, \, x,\, 0) + \mathbb{E} \left[
\int_0^T \frac{\partial^2 F}{\partial \sigma \, \partial x}(\sigma_0,\,
S_s,\, s) \; dS_s \right] \right\} \; \mathcal{A} [\sigma] (\sigma_0) \\ 
\scriptstyle & \scriptstyle & \scriptstyle \\ 
&  & \displaystyle + \frac{1}{2} \left\{ \frac{\partial^2F}{\partial \sigma^2%
}(\sigma_0,\, x,\, 0) + \mathbb{E} \left[\int_0^T \frac{\partial^3 F}{%
\partial \sigma^2 \, \partial x}(\sigma_0,\, S_s, \, s) \; dS_s\right]
\right\} \; \Gamma[\sigma](\sigma_0) \\ 
\scriptstyle & \scriptstyle & \scriptstyle \\ 
\displaystyle \Gamma \left[\mathbb{E}\left[ P\&L \right] \right] & = & %
\displaystyle \left\{ \frac{\partial F}{\partial \sigma}(\sigma_0, \, x, \,
0) + \mathbb{E} \left[ \int_0^T \frac{\partial^2F}{\partial \sigma \;
\partial x}(\sigma_0,\, S_s,\, s) \, dS_s \right] \right\}^2 \; \Gamma[\sigma%
](\sigma_0)%
\end{array}%
\end{equation}

These values represent the inaccuracies that the trader knows to be present
in his estimates.

We can give the following interpretation to this error structure:

\begin{itemize}
\item the bias in the $P \& L$ process represents a deviation in security
prices asked by the trader to the buyer.

\item the variance of the $P \& L$ process naturally generates  a bid/ask
spread on security prices. The width of the bid-ask  spread depends both on
the traders' risk aversion and on the perceived uncertainty on volatility.
\end{itemize}

As a consequence of the presence of the error structure, the price of a
security is thus not unique\footnote{%
When, as in the classical B-S formulation, prices are unique, risk-neutral
arguments can be formulated in order to solve the partial differential
equations which rule the pricing of assets. In this sense, at each instant
in time, price can be represented through a Dirac distribution.}, but it can
be represented, at each instant in time, as a distribution, whose
characteristics depend on the parameters which characterize the error
structure\footnote{%
As pointed out in Scotti\ (2007), PBS\ model induces market incompleteness.}.

Therefore, we have shown that the trader must modify his prices in order to
take into account the two previous effects, namely the variance and the bias
on his expected profit and loss process. Thus, he fixes a supportable risk
probability $\alpha < 0.5$ and accepts to buy the option at a certain price

\begin{equation*}
(\text{Bid Premium}) = (\text{BS Premium}) + \epsilon \; \mathcal{A} \left[%
\mathbb{E}[P \& L] \right] + \sqrt{\epsilon \; \Gamma \left[\mathbb{E}[P \&
L]\right]}\; \mathcal{N}_{\alpha} 
\end{equation*}

where $\mathcal{N}_{\alpha}$ is the $\alpha$-quantile of the reduced normal
law. Analogously, the trader accepts to sell the option at the price

\begin{equation*}
(\text{Ask Premium}) = (\text{BS Premium}) + \epsilon \; A \left[\mathbb{E}%
[P \& L]\right] + \sqrt{\epsilon \; \Gamma \left[\mathbb{E}[P \& L]\right]}
\; \mathcal{N}_{1-\alpha} 
\end{equation*}

Since $\mathcal{N}_{\alpha} + \mathcal{N}_{1- \alpha}= 0$; the mid-premium is

\begin{equation}  \label{rel:mid-price}
(\text{Mid Premium}) = (\text{BS Premium}) + \epsilon \; A \left[\mathbb{E}%
[P \& L]\right]
\end{equation}

and the bid-ask spread is

\begin{equation}  \label{rel:bid-ask}
\text{Bid-Ask spread } = 2 \sqrt{\epsilon \; \Gamma \left[\mathbb{E}[P \& L]%
\right]}\; \mathcal{N}_{\alpha}
\end{equation}

\subsection{European Call options}

We now focus our attention on European call options and we study the bias
and its derivatives in order to derive some sufficient conditions for the
presence of a smiled behaviour on implied volatility. We know the premium of
a call option (see Lamberton et al. \cite{bib:Lamberton_Lapeyre}) with
strike $K$, spot price $x$, volatility $\sigma_0$ and maturity $T$, and we
know its hedging strategy in the usual Black Scholes setting:

\begin{eqnarray*}
F(\sigma_0, \, x, \, 0) & = & x\mathcal{N} (d_1) - K \mathcal{N}(d_2) \\
\text{Delta} & = & \frac{\partial F}{\partial x}(\sigma_0, \, x, \, 0) = 
\mathcal{N} (d_1) \\
\text{where} & & d_1 = \frac{\ln x - \ln K + \frac{\sigma_0^2}{2} T }{%
\sigma_0 \sqrt{T} } \; \; \text{ and } \; \; d_2 = d_1 - \sigma_0 \sqrt{T}.
\end{eqnarray*}

The following results are classical (see \cite{bib:Lamberton_Lapeyre}):

\begin{eqnarray}
\frac{\partial F}{\partial \sigma_0}(\sigma_0, \, x, \, 0) & = & x\sqrt{T} 
\frac{e^{-\frac{1}{2} d_1^2}}{\sqrt{2 \pi}}  \notag \\
\frac{\partial^2F}{\partial \sigma_0^2}(\sigma_0, \, x, \, 0) & = & \frac{x 
\sqrt{T}}{\sigma_0} \frac{e^{-\frac{1}{2} d_1^2}}{\sqrt{2 \pi}} d_1 d_2
\label{eqn_deriv_K2} \\
\frac{\partial^2F}{\partial K^2}(\sigma_0, \, x, \, 0) & = & \frac{x}{K^2
\sigma_0 \sqrt{T}} \frac{e^{-\frac{1}{2} d_1^2}}{\sqrt{2 \pi}}.  \notag
\end{eqnarray}
and we can easily prove that: 
\begin{eqnarray}
\frac{\partial^2 F}{\partial \sigma \, \partial x}(\sigma_0,\, S_s,\, s) & =
& - \frac{1}{\sqrt{2\, \pi} \; \sigma_0} \; d_2(S_s, \, s) \; e^{-\frac{1}{2}
\, d_1^2(S_s, \, s)}  \notag \\
\frac{\partial^3 F}{\partial \sigma^2 \, \partial x}(\sigma_0, \, S_s, \, s)
& = & \frac{d_1(S_s, \, s) + d_2(S_s, \, s) - d_1(S_s, \, s)\, d_2^2(S_s, \,
s)}{\sqrt{2 \, \pi} \; \sigma_0^2} \; e^{-\frac{1}{2} \, d_1^2(S_s, \, s)} \\
\text{where} & & d_1(S_s, \, s) = \frac{\ln S_s - \ln K + \frac{\sigma_0^2}{2%
} (T-s) }{\sigma_0\, \sqrt{T-s} }  \notag \\
\text{and} & & d_2(S_s, \, s) = d_1(S_s, \, s) - \sigma_0\, \sqrt{T-s} 
\notag
\end{eqnarray}

We apply the Perturbative Black Scholes model to find the corrections it
imposes on the expected profit and loss process for a trader who is hedging
a short position on a plain vanilla European call option. Then the bias on
the call premium is given by two terms. The first one is the bias when $\mu
= 0$. This case is accurately studied in \cite{bib:Scotti-PBS}.

\begin{equation}  \label{eqn:first-corr}
A_{\mu=0}[C]|_{\sigma= \sigma_0} = x \; \frac{e^{-\frac{1}{2} d_1^2}}{\sqrt{%
2 \, \pi}} \, \left\{ A\left[\sigma \, \sqrt{T}\right]|_{\sigma= \sigma_0} + 
\frac{d_1 \, d_2}{2 \, \sigma_0 \, \sqrt{T}} \; \Gamma \left[\sigma \, \sqrt{%
T} \right]|_{\sigma= \sigma_0} \right\}
\end{equation}

Now, if we assume that the drift term of the stock price process is non
zero, this correction is not sufficient in order to hedge the position
correctly. We have to study another term, which is the correction when $\mu
\neq 0$. While when $\mu = 0$ the stochastic integrals in equation (\ref%
{relation-bias-variance}) are martingales, if $\mu>0$ we have to evaluate
their expectations:

\begin{eqnarray*}
A_{\text{correction}}[C]|_{\sigma= \sigma_0} & = & \mathbb{E} \left[
\int_0^T \frac{\partial^2 F}{\partial \sigma \, \partial x}(\sigma_0,\,
S_s,\, s) \; dS_s \right] \; \mathcal{A} [\sigma] (\sigma_0) \\
& & + \frac{1}{2} \mathbb{E} \left[\int_0^T \frac{\partial^3 F}{\partial
\sigma^2 \, \partial x}(\sigma_0,\, S_s, \, s) \; dS_s\right] \; \Gamma[%
\sigma](\sigma_0)
\end{eqnarray*}

In appendix \ref{app:calcoli}, we compute the two integrals and we find

\begin{equation}  \label{eqn:second-corr}
\begin{array}{rcl}
\displaystyle A_{\text{correction}}[C]|_{\sigma= \sigma_0} & = & %
\displaystyle K \; \left\{ \frac{\mathcal{N}\left( d_2 + \mathcal{L} \right)
- \mathcal{N}\left( d_2 \right) }{\mathcal{L}} -\frac{1}{\sqrt{2 \, \pi}}
e^{-\frac{1}{2}d_2^2} \right\} \; \left. A\left[\sigma \, \sqrt{T}\right]%
\right|_{\sigma= \sigma_0} \\ 
\scriptscriptstyle & \scriptscriptstyle & \scriptscriptstyle \\ 
&  & \displaystyle - \frac{K}{\sigma_0 \, \sqrt{T}} \left\{ \left[ \sigma_0
\, \sqrt{T} \, \left[ \frac{3}{\mathcal{L}^2} + \left(1+ \frac{d_2}{ 
\mathcal{L} }\right)^2 \right] - \, \frac{8}{\mathcal{L} } - 6 \frac{d_2}{%
\mathcal{L}^2 } \right] \left[ \mathcal{N} \left(d_2 + \mathcal{L} \right) - 
\mathcal{N} \left(d_2 \right) \right] \right. \\ 
\scriptscriptstyle & \scriptscriptstyle & \scriptscriptstyle \\ 
&  & \displaystyle + \frac{1}{\sqrt{2 \, \pi}} \left[ d_2^2 +4 -2 \, \frac{%
d_2}{\mathcal{L}} + \frac{8}{\mathcal{L}^2 } + \sigma_0 \, \sqrt{T} \left(
d_2 - \frac{ 4}{\mathcal{L} } - \frac{ d_2}{\mathcal{L}^2} \right) \right]
e^{-\frac{1}{2}\, d_2^2} \\ 
\scriptscriptstyle & \scriptscriptstyle & \scriptscriptstyle \\ 
&  & \displaystyle \left. + \frac{1}{\sqrt{2 \, \pi}} \left[ \frac{\sigma_0
\, \sqrt{T}}{\mathcal{L}} \, \left(1 + \frac{d_2}{ \mathcal{L}}\right) - 
\frac{8}{\mathcal{L}^2}\right] \, e^{-\frac{1}{2} \left( d_2 + \mathcal{L}
\right)^2} \right\}\; \left. \Gamma \left[\sigma \, \sqrt{T} \right]%
\right|_{\sigma= \sigma_0}%
\end{array}%
\end{equation}
where $\mathcal{L}$ is the cumulated risk premium:

\begin{equation}
\mathcal{L} = \lambda \, \sqrt{T} = \frac{\mu \, T}{\sigma_0 \, \sqrt{T}}
\end{equation}

\begin{remarque}
We remark that the first correction (\ref{eqn:first-corr}) derives from the
bias of the option price. It is then an uncertainty coming from the error on
estimating the value of volatility.

The second correction, (\ref{eqn:second-corr}) is a consequence of the
presence of a bias on the strategy, which introduces uncertainty on the
hedging procedure also.
\end{remarque}

It is easy to compute the value of the variance term of the error structure
for a call option:

\begin{equation}
\Gamma[Call] = \left\{\frac{x}{\sqrt{2 \, \pi}} e^{-\frac{1}{2} d_1^2} + K
\; \left[ \frac{\mathcal{N}\left( d_2 + \mathcal{L} \right) - \mathcal{N}%
\left( d_2 \right) }{\mathcal{L}} -\frac{1}{\sqrt{2 \, \pi}} e^{-\frac{1}{2}%
d_2^2} \right] \right\}^2 \; \Gamma[\sigma \sqrt{T}]
\end{equation}

We assume that the ask price is then simply the mid price increased by a
standard deviation, symmetrically the bid price is the mid price decreased
by a standard deviation. The spread is simply given by

\begin{equation}
\text{Bid-Ask spread}\, [Call] = 2\, \sqrt{\epsilon \;\Gamma[Call]}
\end{equation}

\section{Numerical Analysis}

In this section we explore the sensitivity of the PBS\ model to some
parameters and, through numerical analysis, we give evidence of the fact
that the model is able to reproduce all the observed behaviours of the
implied volatility curve. The existence of closed form solutions to the
pricing formulas allows us to make some comparative static exercises in
order to analyze the dependence of the implied volatility curve on time
horizon and, above all, on the drift term.

\subsection{Parameter sensitivity}

First of all, we point out that the corrections obtained with respect to
Black and Scholes prices depend on the choice of the parameters A and $%
\Gamma$ and on the magnitude of the $\epsilon$ term\footnote{ 
Since we have three parameters, a possible way to calibrate the model to the
market behaviour is by using instruments which price variance (e.g.
variances swaps). Such derivatives permit to find an implicit link between
the bias and the variance term; by fixing an epsilon, it is possible then to
calibrate the model on just one parameter, see Scotti \cite{bib:Scotti-VS}.
Another possibility is instead to fix an arbitrary epsilon small and use the
implied spread to calibrate the two coefficients A and $\Gamma$.}. A captures
the bias introduced on the profit and loss function, while $\Gamma \ $is a
variance term. $\varepsilon $ is just a scale factor. It must be small
enough to make the higher order expansion terms be negligible. Let us
analyze the sensitivity of the volatility implied by the PBS model to the
choice of these parameters. Thus, we fix a value for $\sigma \ $and the
other real world parameters and we let the coefficients of the error
structure vary.

\begin{figure}[h!!]
  \begin{center}
    \epsfxsize=16.5cm
    \epsfbox{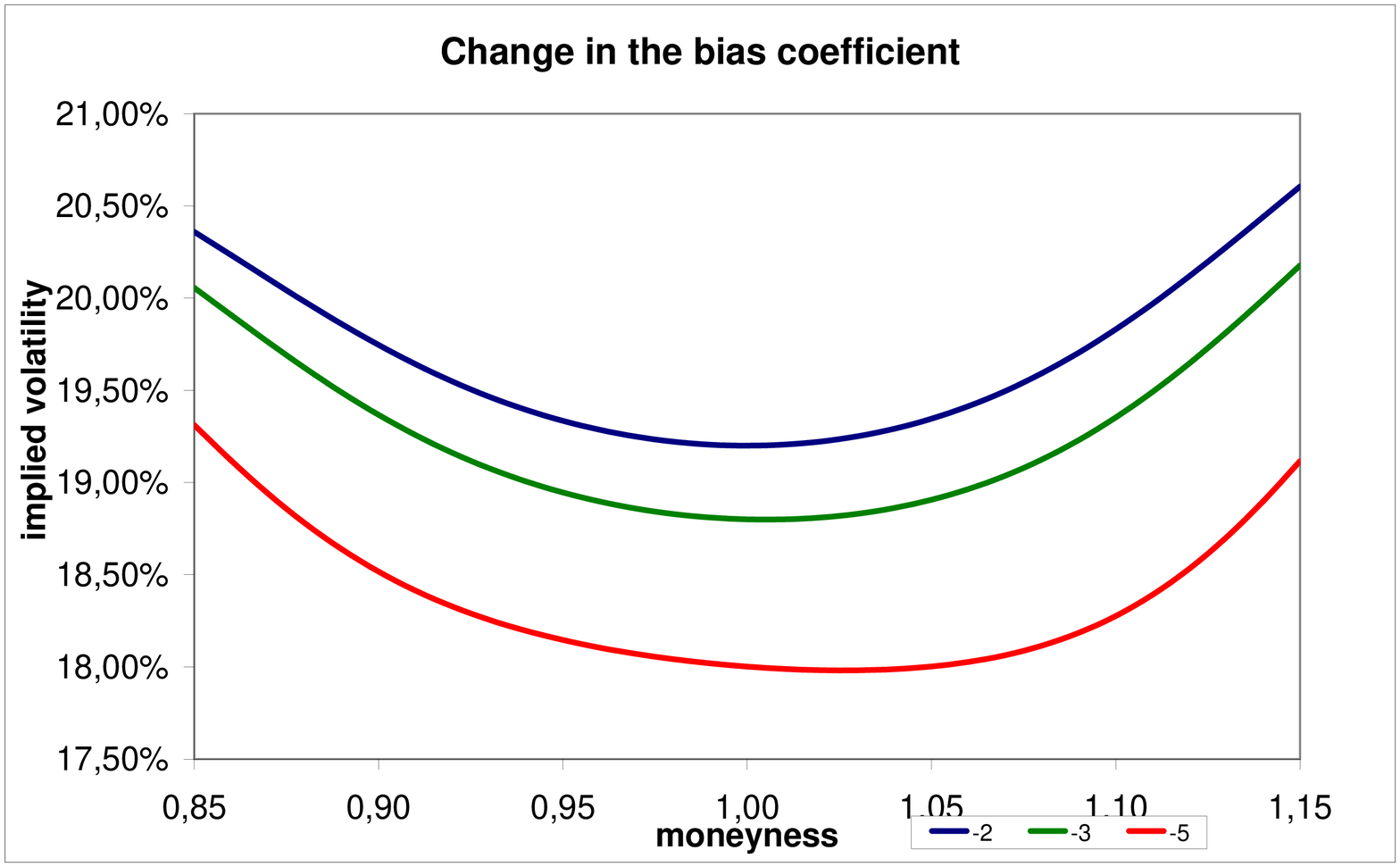}
    \vspace{-1.5cm}
    \caption{Implied volatilities curves
      depending on $A[\sigma]$.}\label{DRIFT-fig-1}
    \vspace{-0.5cm}
    \epsfxsize=16.5cm
    \epsfbox{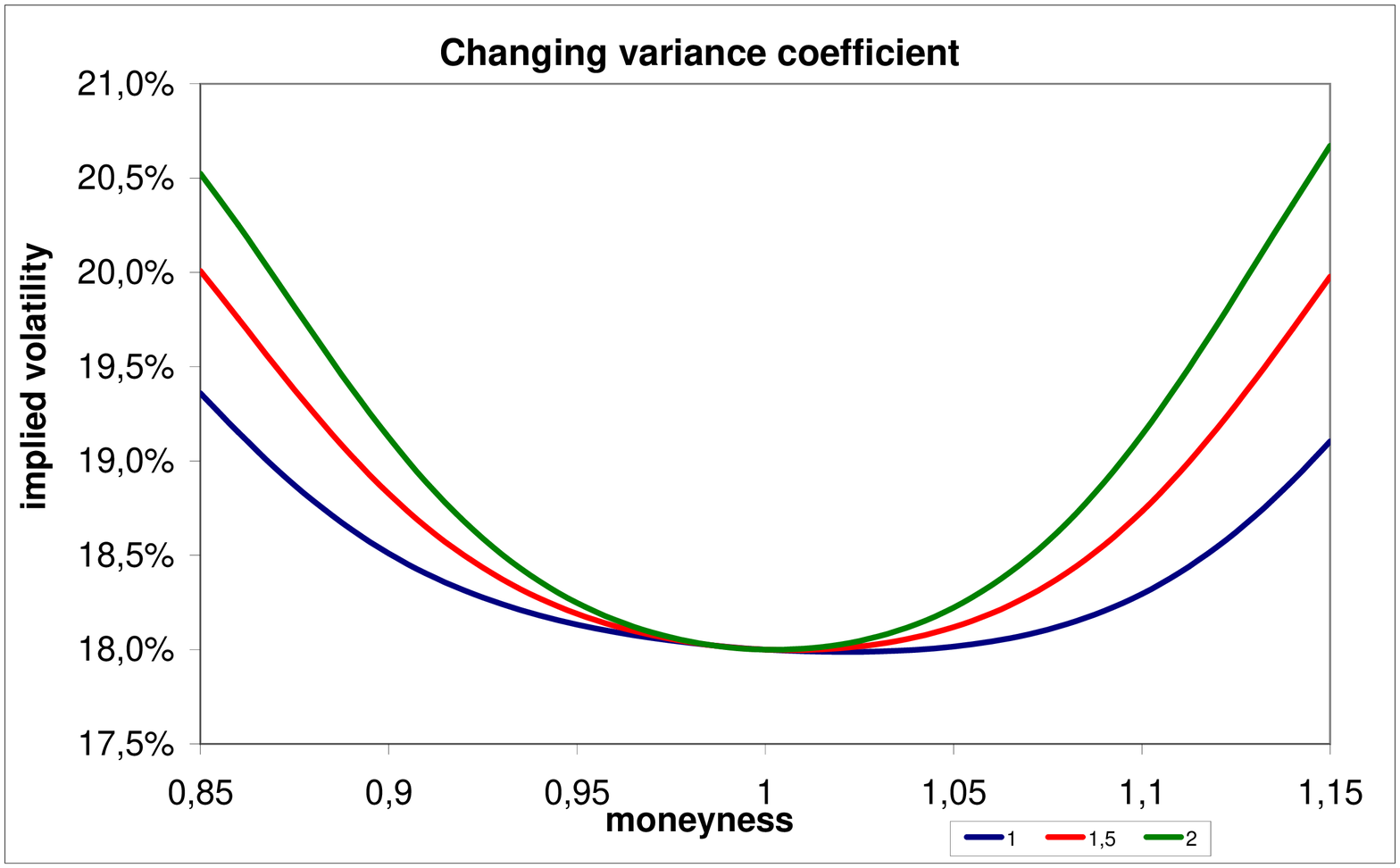}
    \vspace{-1.5cm}
    \caption{Implied volatilities curves
      depending on $\Gamma[\sigma]$.}\label{DRIFT-fig-2}
  \end{center}
\end{figure}

Figure \ref{DRIFT-fig-1} shows the effect of an increase in the
absolute value of the coefficient of the bias term. As we will
explain below, there are good reasons for considering a negative
bias. Then, the lower the coefficient, the more the curve shifts
downwards and the point of minimum variance to the right. Increasing
the value of the coefficient of the variance term, instead, clearly
``opens'' up the smile, which becomes more pronounced. Moreover, as
can be seen in figure \ref{DRIFT-fig-2}, a higher coefficient is
associated with a more pronounced skew effect which makes the
implied volatility higher for out-of-the money options, compared to
in-the-money ones.

\begin{figure}[h!!]
  \begin{center}
    \epsfxsize=16.5cm
    \epsfbox{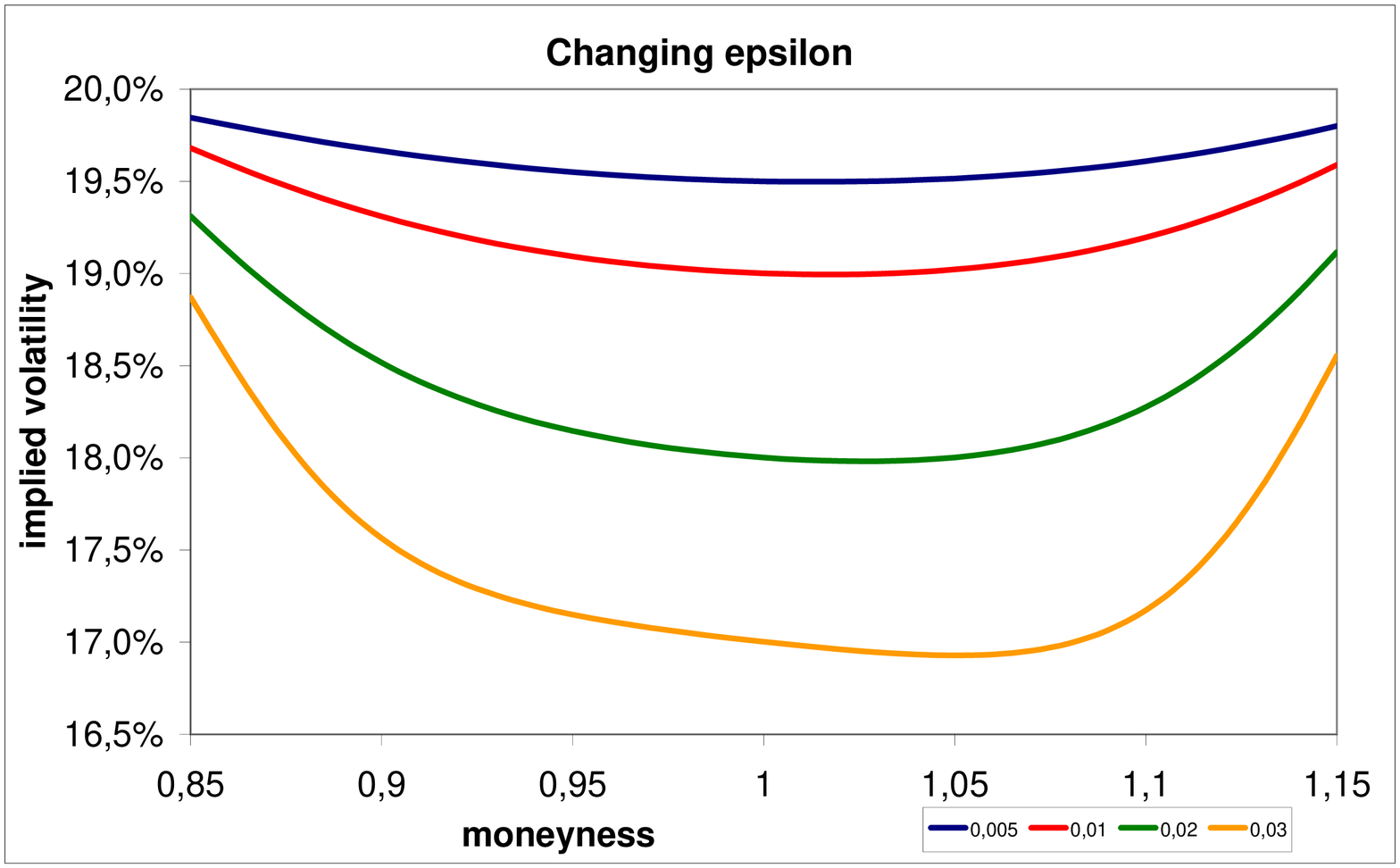}
    \vspace{-1.5cm}
    \caption{Implied volatilities curves
      depending on $\epsilon$.}\label{DRIFT-fig-3}
    \vspace{-0.5cm}
    \epsfxsize=16.5cm
    \epsfbox{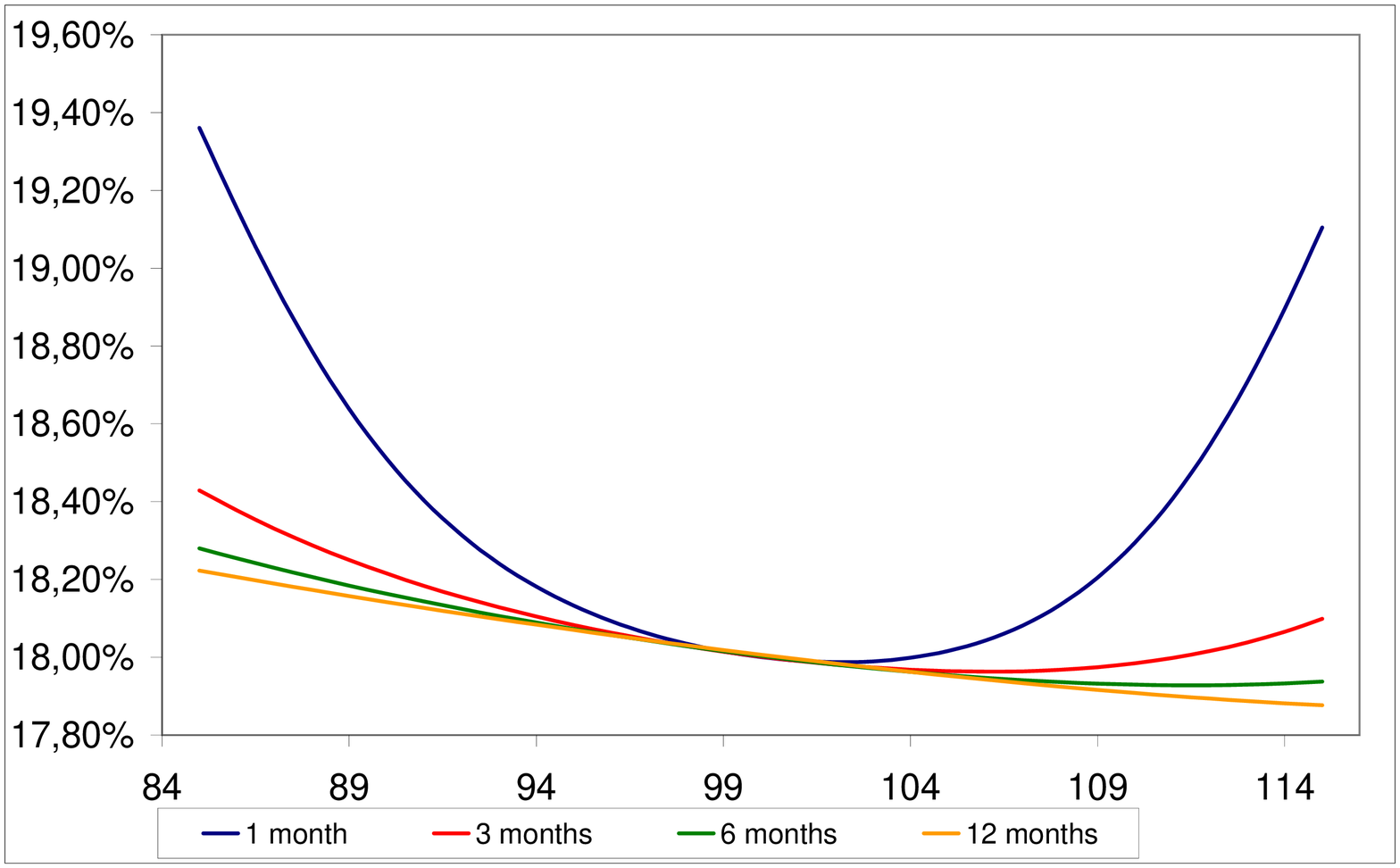}
    \vspace{-1.5cm}
    \caption{Skewed implied volatility
      depending on maturity.}\label{DRIFT-fig-4}
  \end{center}
\end{figure}

A change in epsilon, instead, combines the two effects described above.
Notice that this parameter must be small enough to justify expansion 
(\ref{eqn:chain-rule}) in our case. A higher epsilon, thus, produces both a
downward shift of the curve and more pronounced smile and skew
effects, see figure \ref{DRIFT-fig-3}.

From now on, we fix the values of these three parameters to make
some comparative static analysis of the parameters that capture the
real world features. We set epsilon to $0.02$ for convenience.

The coefficient on the variance term is set, by a normalization argument%
\footnote{Obviously, this coefficient can not be negative.  As shown
in Figure \ref{DRIFT-fig-2}, higher values of this  coefficient lead
to more pronounced smiles.}, to

\begin{equation}
\Gamma \lbrack \sigma ]_{\sigma _{0}}=\sigma _{0}^{\ 2}
\end{equation}

implying that

\begin{equation}
Variance=\varepsilon \sigma _{0}^{\ 2}
\end{equation}

The bias coefficient is instead set to

\begin{equation}
A[\sigma ]|_{\sigma _{0}}=-5\sigma _{0}
\end{equation}

leading to

\begin{equation}
Bias=-5\varepsilon \sigma _{0}
\end{equation}

This last choice is made in order to reproduce a precautionary effect. The
hypothesis we make is that for some reason, the trader believes he
overestimated volatility in his procedure. This feeling can be justified by
two reasons, one mathematical and one economic.

The mathematical explanation lies in the analysis of the usual formula used
to estimate historical volatility under the hypothesis of lognormality of
stock prices. Since it is a concave function, it is more likely that the
approximated value found by the estimation procedure is an overestimation of
the true one.

The economic explanation lies in the way volatility is usually described in
models. Markets are opened for 8 hours a day only. However, the flow of
information does not stop when markets are closed: variability accumulates
even if securities are not traded. Then, in almost every pricing model,
volatility is described as a continuous process. This is of course a
simplification, but seems nevertheless reasonable. However, it has been
shown by some authors, see Stoll and Whaley \cite{bib:Stoll_Whaley} that
overnight volatility is consistently lower than intra day one. Hence, it is
straightforward to believe that usual models overestimate volatility.

Let us consider the PBS model prediction on a one-month European call option
with the parameters we set above.

First, we keep the risk premium measure fixed and we analyze how implied
volatility changes with different maturities. Our finding is that we obtain
curves which are flatter as long as the option time horizon becomes longer.
This behaviour is consistent with empirical evidence on almost every
derivative market, see Hagan et al. \cite{bib:SABR}. Figure 4 shows that the
implied volatility curve is skewed to the left for each option; the point of
minimum variability shifts towards higher strikes for longer time horizons.

As we showed in the previous section, the use of our perturbative approach
implies that option prices and, thus, implied volatilities are affected by
changes in the risk premium. If we let $\lambda \ $, the risk premium we
defined previously, change and we fix the parameters that characterize the
error structure, we can observe and analyze this sensitivity.

\begin{figure}[h!!]
  \begin{center}
    \epsfxsize=16.5cm
    \epsfbox{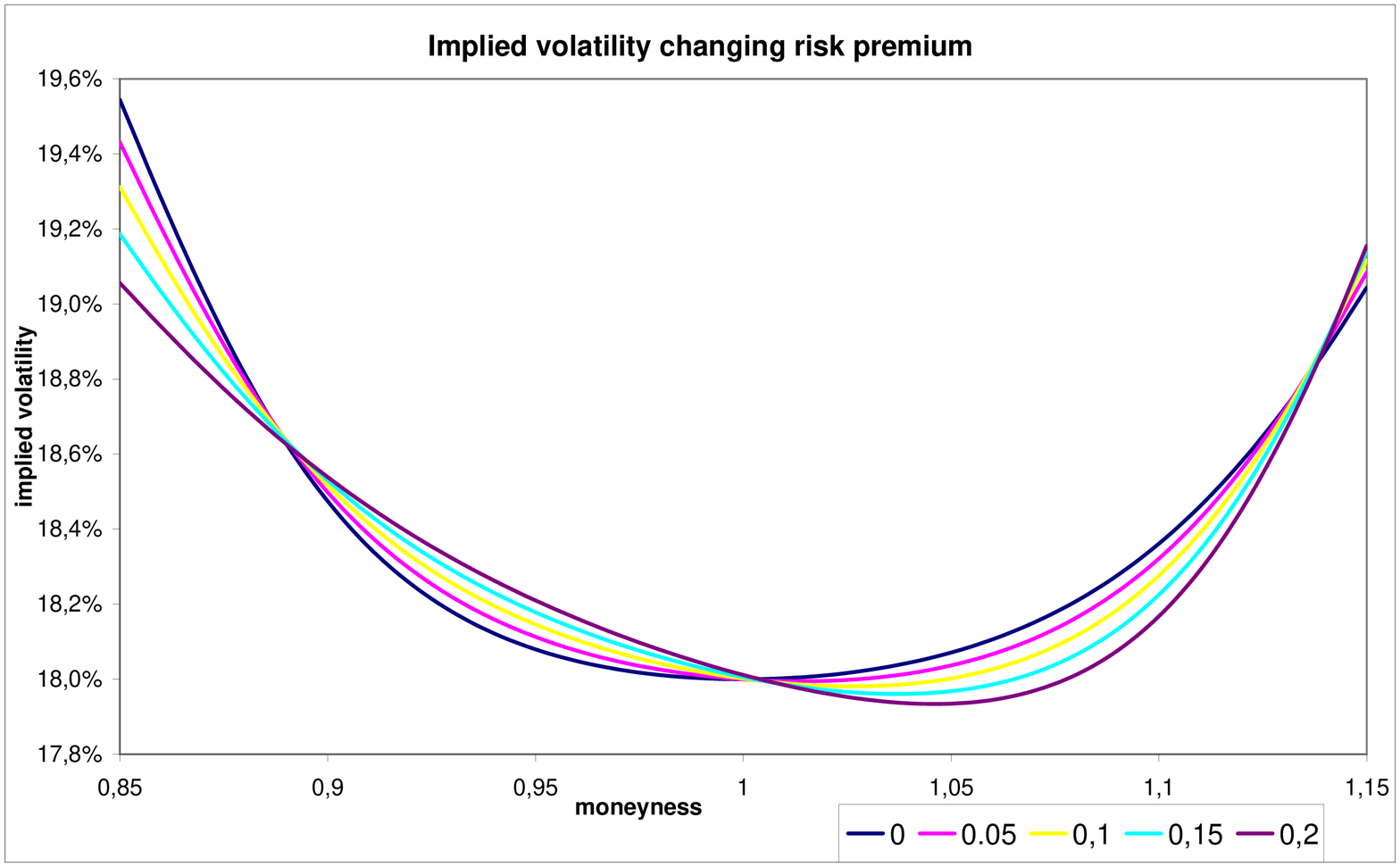}
    \vspace{-1.5cm}
    \caption{Implied volatility
      depending on $\lambda$.}\label{DRIFT-fig-5}
 \vspace{-0.5cm}
    \epsfxsize=16.5cm
    \epsfbox{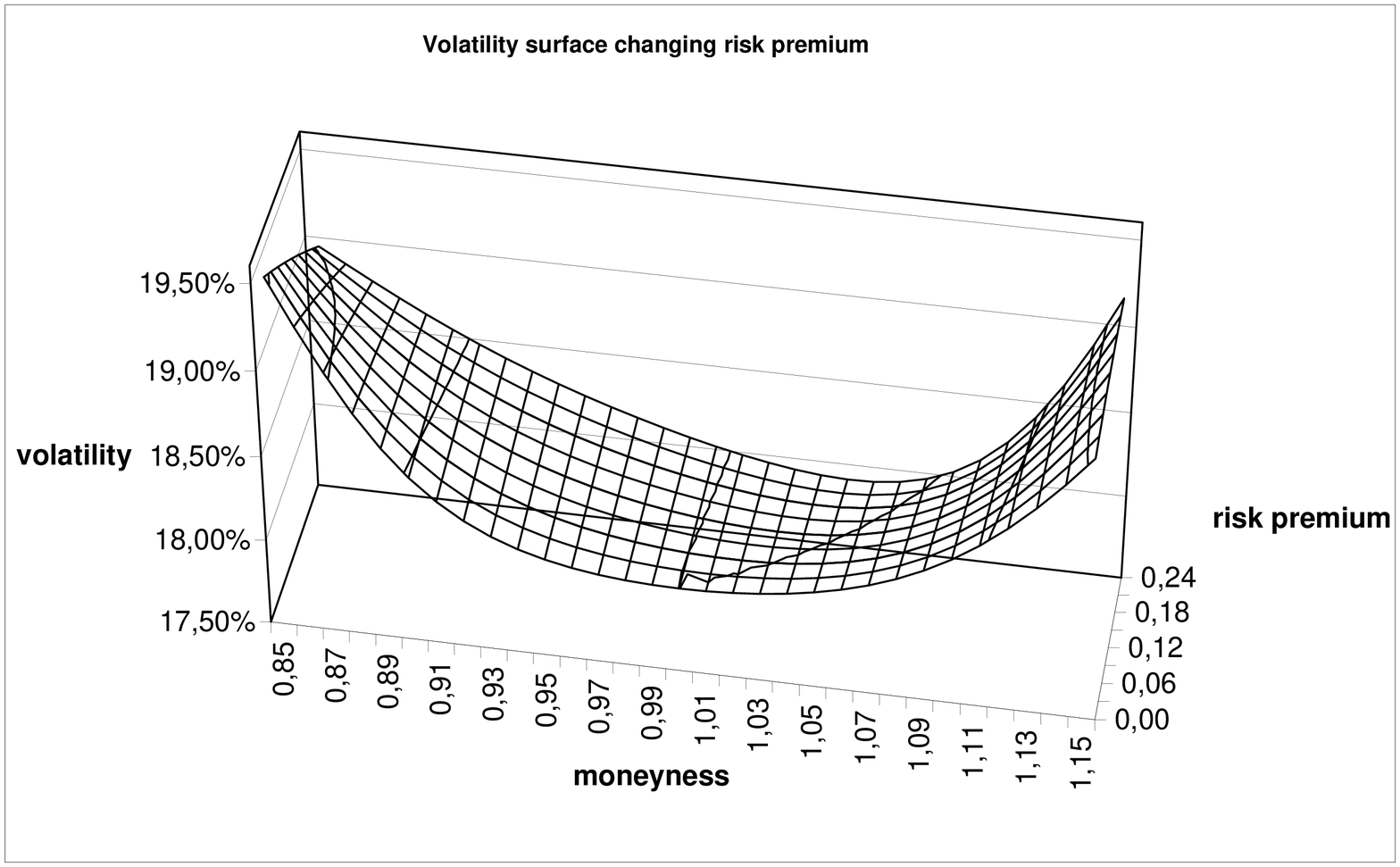}
    \vspace{-1.5cm}
    \caption{Implied volatility
      depending on $\lambda$.}\label{DRIFT-fig-6}
  \end{center}
\end{figure}

Figures \ref{DRIFT-fig-5} and \ref{DRIFT-fig-6} show the behavior of
the implied volatility curve on a $1$ month European call option for
an expected excess return on stock term that ranges from $0$ to
$0.2$. The curve evidently shifts to the right side of the graph as
the risk premium term increases. With almost every value up to
$0.2$, i.e $\lambda=1$, there is a skew effect towards lower
maturities. The lower the risk premium, the higher is the value of
implied volatility for deep in the money options and the lower for
options which are far out of the money. As the value of $\mu$
increases the curve appears to become steeper on the right side. In
particular, for this parameter choice, for a very high risk premium,
there is a slight tendency to change the skew direction\footnote{
For $\mu = 0.2$ the implied volatility at moneyness $1.15$ is
slightly higher than at $0.85$. Unreported simulations show that
this behavior is common to every choice of parameter. This could
suggest that in periods of high risk premia, volatility should tend
to be higher for out of the money options.}. Curves cross
approximately at the money, around $102$.

These findings are consistent with those obtained by other stochastic
volatility models, such as the SABR\ one (Hagan et al. \cite{bib:SABR}). The
authors of that model, fitting it on prices of Eurodollar options, obtained
those behaviours of the implied volatility curve, under the hypothesis that
asset prices and volatilities are correlated.

\subsection{Spread Analysis}

Up to now, we have considered the mid price only. As pointed out
before, the perturbative approach used by the PBS model can
naturally generate a spread on prices and volatilities.

The spread on implied volatility is then obtained by inversion of the
pricing formula.

\begin{figure}[h!!]
  \begin{center}
    \epsfxsize=16.5cm
    \epsfbox{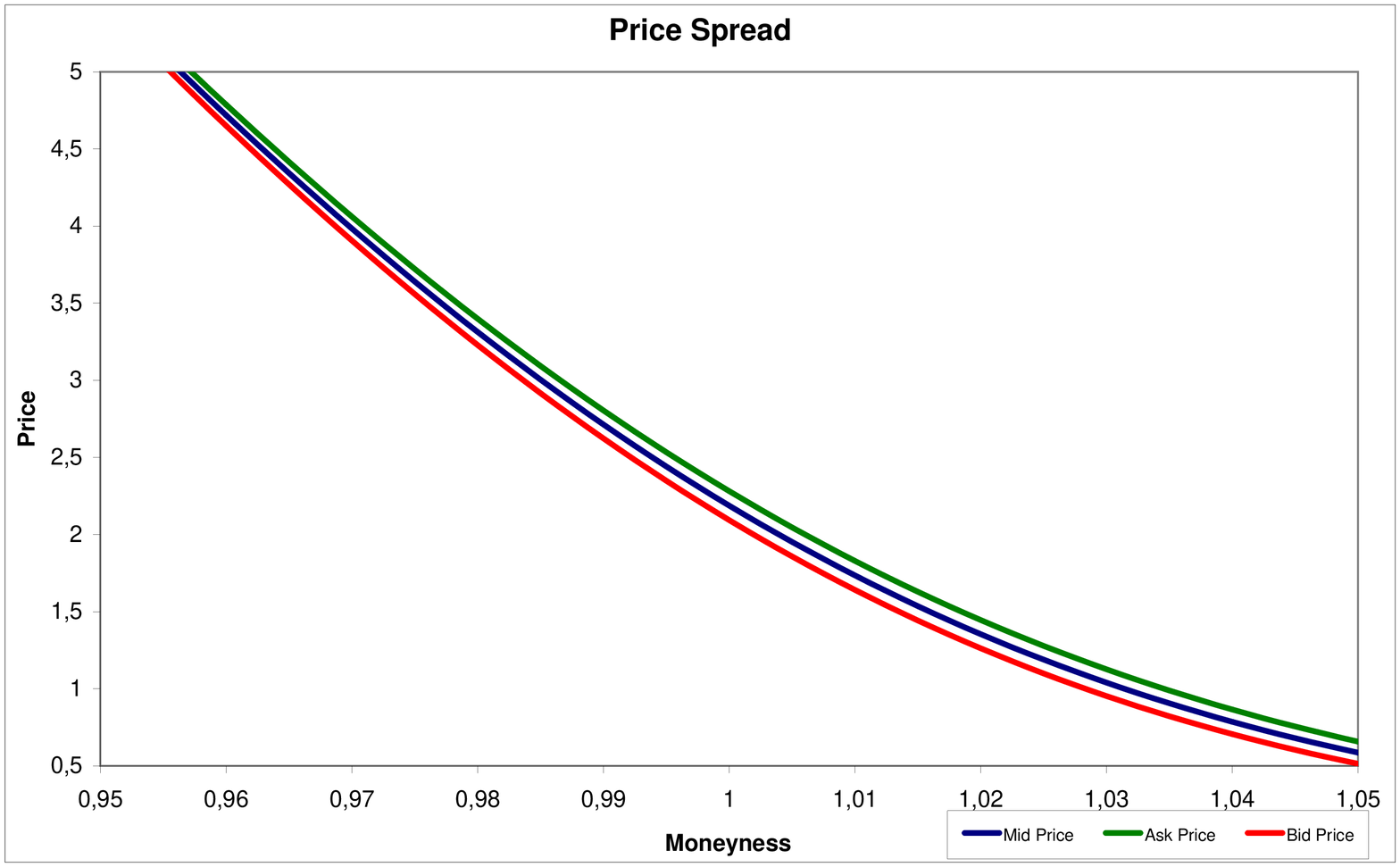}
    \vspace{-1.5cm}
    \caption{Price spread.}\label{DRIFT-fig-7}
    \vspace{-0.5cm}
    \epsfxsize=16.5cm
    \epsfbox{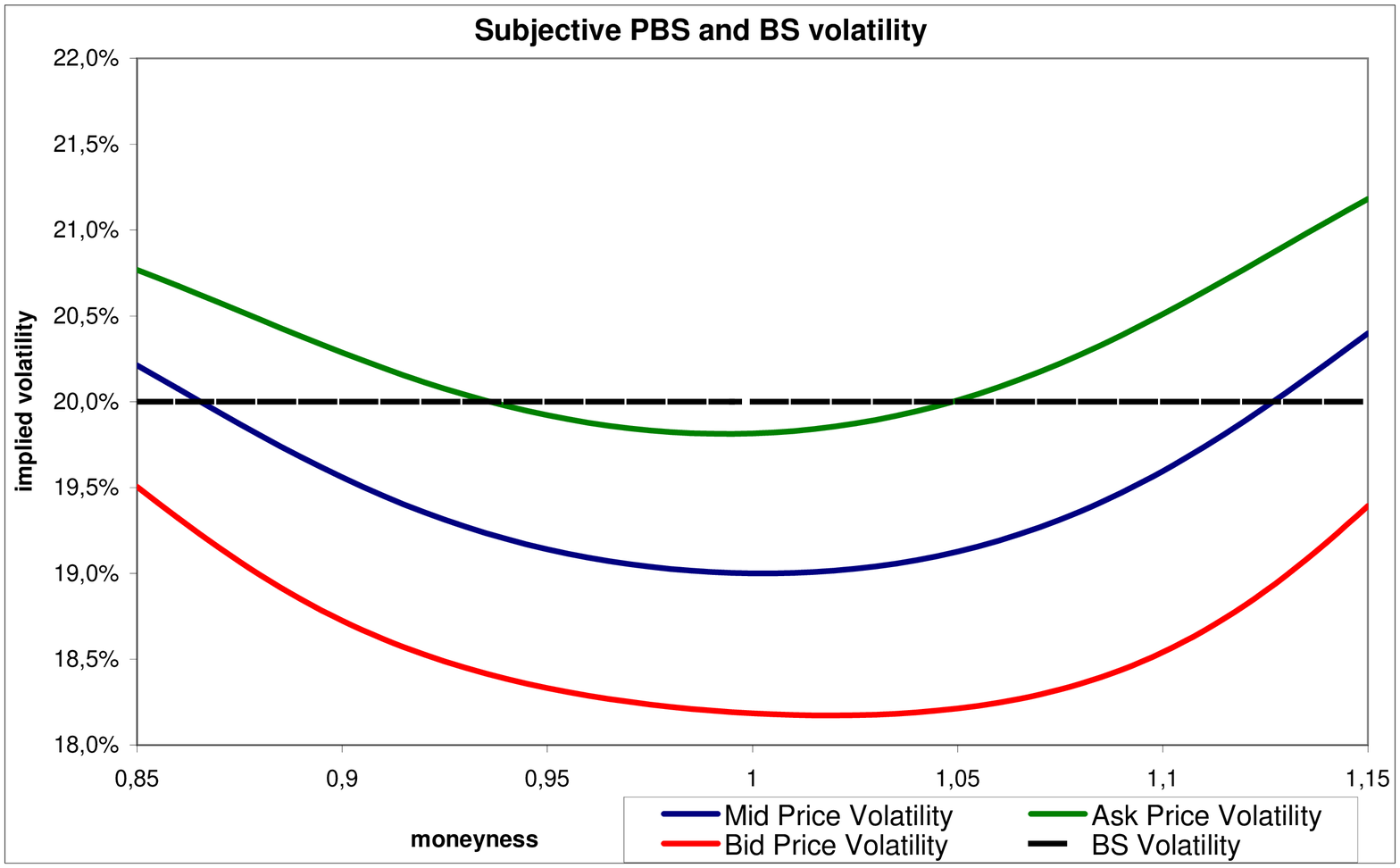}
    \vspace{-1.5cm}
    \caption{Percentual price spread.}\label{DRIFT-fig-8}
  \end{center}
\end{figure}

\begin{figure}[h!!]
  \begin{center}
    \epsfxsize=16.5cm
    \epsfbox{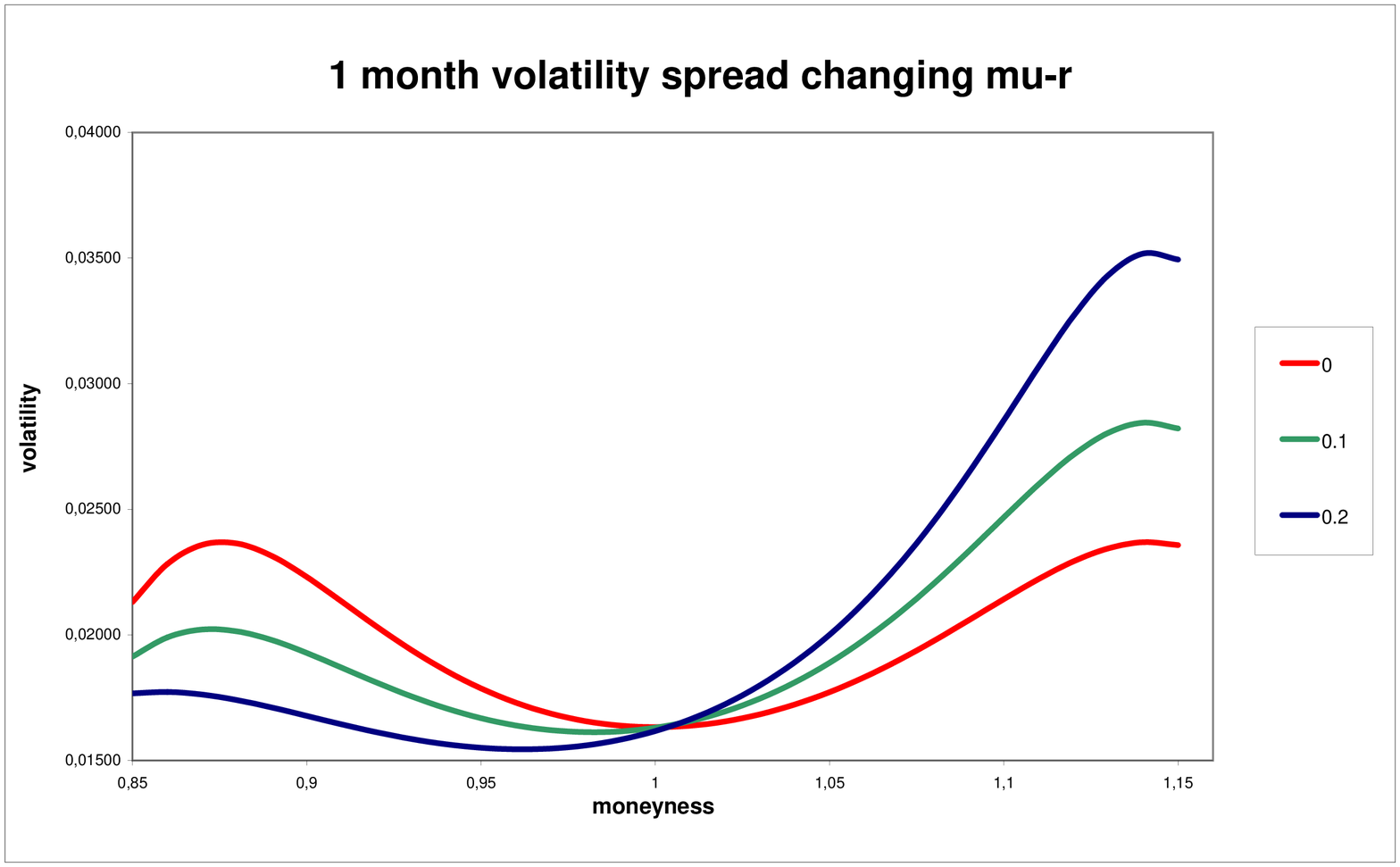}
    \vspace{-1.5cm}
    \caption{Volatility spread
      depending on $\lambda$.}\label{DRIFT-fig-9}
    \vspace{-0.5cm}
    \epsfxsize=16.5cm
    \epsfbox{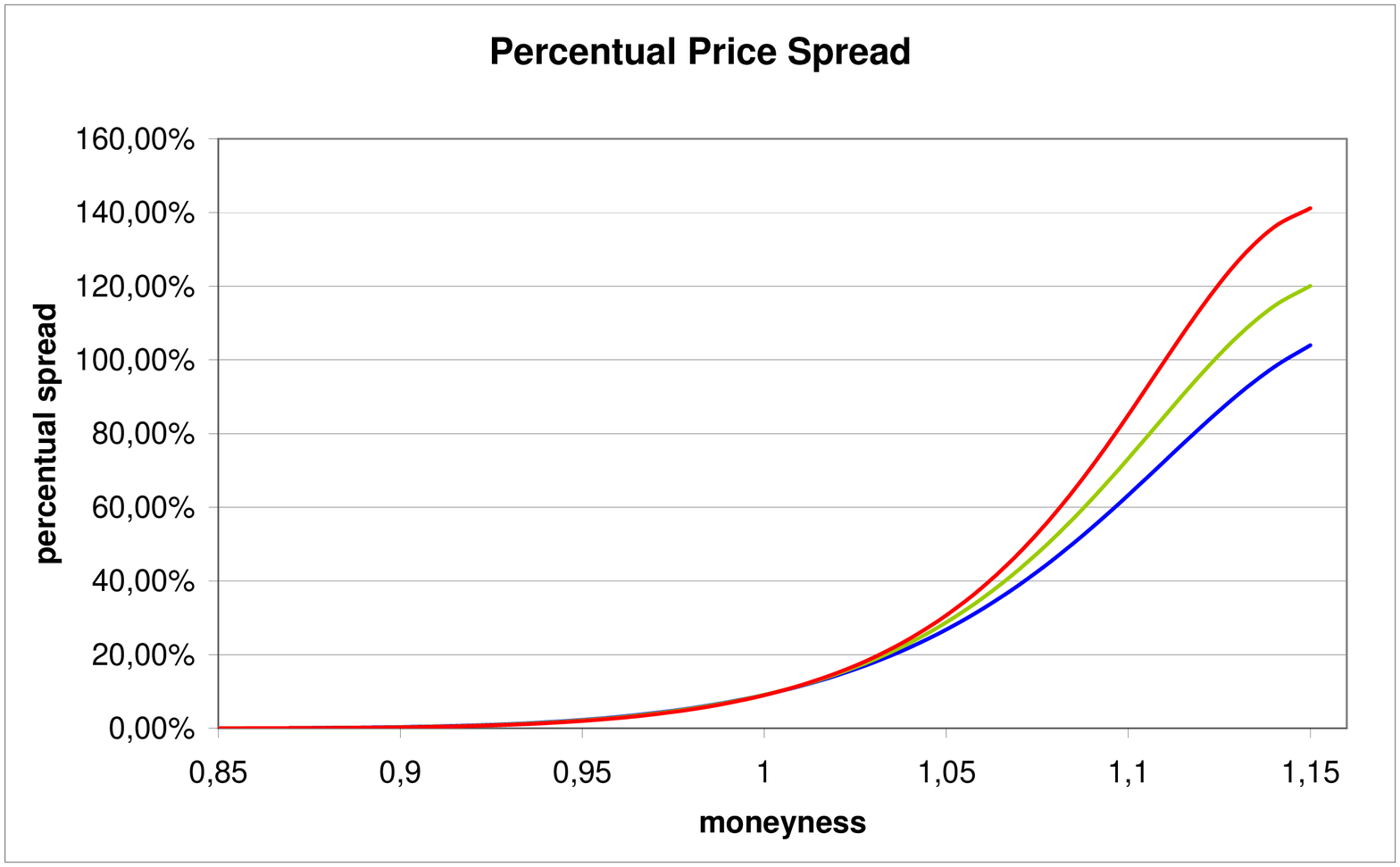}
    \vspace{-1.5cm}
    \caption{Implied volatility in PBS model
      with and without drift.}\label{DRIFT-fig-10}
\end{center}
\end{figure}

For the same set of parameters described above, we can thus analyze
the effect of changing the drift term on a theoretical bid-ask
spread. Figures \ref{DRIFT-fig-7} and \ref{DRIFT-fig-8} give an
example of price and volatility spread behavior for a chosen value
of $\mu$. It is straightforward to notice that higher prices imply
higher volatility. Figure \ref{DRIFT-fig-9} shows the magnitude of
the spread on implied volatility\footnote{Notice that the analysis
of the relative spread leads to the same conclusions.} for three
different values of $\mu$. For low strikes, the spread is higher
when there is no risk premium; it reaches a minimum around the
money, then it starts increasing. For out of the money options, the
behavior is reversed: the spread is higher the higher the risk
premium. The main difference we find with the standard $\mu=0$
setting is that the spread has no longer its point of minimum
variance around the money. The variance spread becomes indeed wider
as the strike increases. As shown in figure \ref{DRIFT-fig-10}, the
relative spread on prices (spread-mid price) is almost zero for in
deep in the money options, then increases sharply with both strike
and risk premium for out of the money calls.

\begin{figure}[h!!!]
  \begin{center}
    \epsfxsize=16.5cm
    \epsfbox{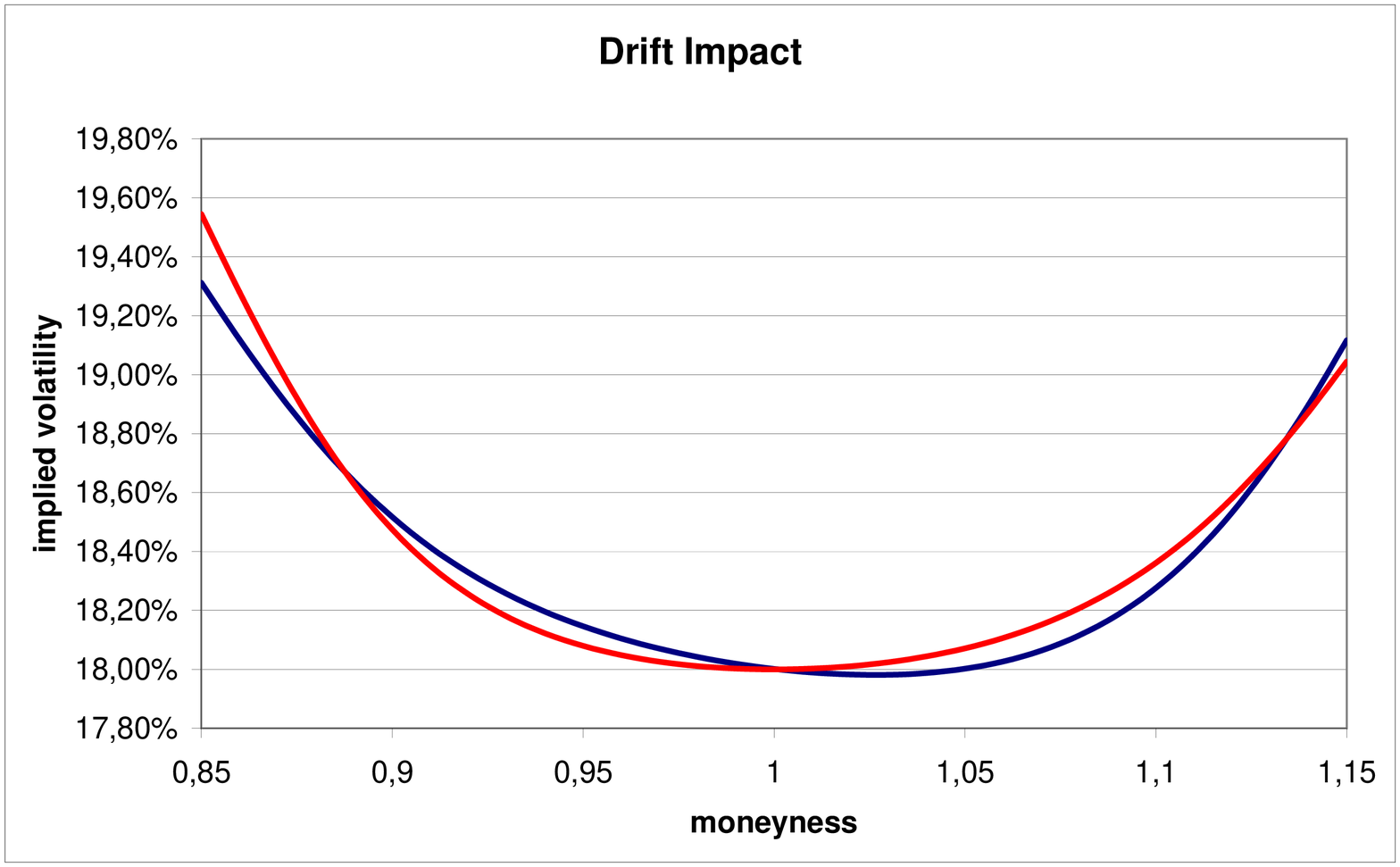}
    \vspace{-1.5cm}
    \caption{Bid, ask, mid implied volatilities
      in PBS model.}\label{DRIFT-fig-11}
  \end{center}
\end{figure}

Let us finally consider directly the effect of the addition of the
correction in equation (\ref{eqn:second-corr}) to the PBS
model implied volatility curve. Figure \ref{DRIFT-fig-11} clearly
shows that the presence of a risk premium skews the implied
volatility curve toward higher strikes.

\newpage

\section{Conclusion}

In this article we have studied the Perturbative Black Scholes model,
introduced in \cite{bib:Scotti-PBS}, when we drop the hypothesis that the
underlying is a martingale under the historical probability. Without
imposing any behaviour of volatility through time, we showed that the
hedging procedure of a trader who estimates it depends on the expected
excess return on stocks.

We then introduced a correction with respect to Scotti \cite{bib:Scotti-PBS}
when the drift term of the diffusive process for the stock price is
different from the risk free rate. We found a closed form solution for the
pricing of a European vanilla call option. This formula depends on the same
parameters of the classical Black Scholes model, i.e. the volatility $%
\sigma_0$ and the parameter $d_2$, on the two parameters of the PBS model,
the variance $\Gamma[\sigma]$ and the bias $A[\sigma]$, which characterize
the error structure of the volatility estimated by the trader. Since the PBS
model induces market incompleteness, pricing formulas depend also on the
cumulated risk premium $\mathcal{L}$, as shown by equation(\ref%
{eqn:second-corr}).

We analyzed how a simple risk aversion argument forces the underlying price
to be a sub-martingale and we studied the dependence of implied volatility
on the parameters of the model. We numerically studied the most case in
which the volatility used by traders is an overestimation of the true value
and we showed that higher risk premia tend to increase the skewness and the
smile of the implied volatility curve, since the distribution of stock
prices at maturity is shifted towards higher values.

We finally found out that the Perturbative Black Scholes model with drift
can reproduce the behaviour of the implied volatility curve after the 1987
crash.

\appendix

\section{Computation}

\label{app:calcoli}

We have to compute 
\begin{equation}  \label{eqn:first-term}
\mathbb{E} \left[ \int_0^T \frac{\partial^2 F}{\partial \sigma \, \partial x}%
(\sigma_0, \, S_s, \, s)\; dS_s \right]
\end{equation}
where $S_t$ follows the Black Scholes diffusion (\ref{eqn:SDE-BS}) and $%
F(\sigma_0, \, S_s, \, s)$ is the price of a call option with strike K,
starting at time s, when the spot value is $S_s$ and the volatility is $%
\sigma_0$.

\begin{eqnarray*}
\mathbb{E} \left[ \int_0^T \frac{\partial^2 F}{\partial \sigma \, \partial x}%
(\sigma_0, \, S_s, \, s) \; dS_s \right] & = & - \frac{ \mu}{\sqrt{2 \, \pi}
\; \sigma_0} \int_0^T \mathbb{E}\left[d_2(S_s, \, s) \; e^{-\frac{1}{2} \,
d_1^2(S_s, \, s)} \; S_s \right] ds \\
& = & - \frac{ \mu \, S_0 }{\sqrt{2\, \pi} \; \sigma_0} \int_0^T
e^{\left(\mu- \frac{1}{2} \sigma^2_0\right)\, s}\; \mathbb{E}\left[ \frac{%
\ln \frac{S_0}{K} + \mu \, s - \frac{1}{2}\sigma_0^2 s + \sigma_0 \, W_s - 
\frac{\sigma_0^2(T-s)}{2} }{\sigma_0 \, \sqrt{T-s}} \right. \\
& & \star \; \left. e^{ -\frac{1}{2} \left\{\frac{\ln \frac{S_0}{K}+ \mu \,
s - \frac{1}{2}\sigma_0^2 s + \sigma_0 \, W_s + \frac{\sigma_0^2(T-s)}{2}}{
\sigma_0 \, \sqrt{T-s}}\right\}^2 } \; e^{\sigma_0 \,W_s} \right] ds \\
& = & - \frac{ \mu S_0}{\sqrt{2 \pi} \, \sigma_0} \; \int_0^T e^{ \left(\mu- 
\frac{1}{2}\sigma^2_0\right)\, s} \int_{\mathbb{R}} \frac{\ln \frac{S_0}{K}
+ \mu \, s - \frac{1}{2}\sigma_0^2\, s + \sigma_0 \, \sqrt{s} \, y - \frac{%
\sigma_0^2\,(T-s)}{2} }{\sigma_0 \, \sqrt{T-s}} \\
& & \star \; e^{ -\frac{1}{2} \left\{\frac{\ln \frac{S_0}{K}+ \mu \, s - 
\frac{1}{2}\sigma_0^2 \, s + \sigma_0 \, \sqrt{s} \, y + \frac{%
\sigma_0^2\,(T-s)}{2}}{ \sigma_0 \, \sqrt{T-s}}\right\}^2} \; e^{\sigma_0 
\sqrt{s} y} \; \frac{e^{-\frac{1}{2}\, y^2}}{\sqrt{2 \pi}} \; dy \; ds \\
& = & - \frac{ \mu \, K}{\sqrt{2\, \pi} \, \sigma_0} \int_0^T \frac{T-s}{T}
\; \left[\frac{\ln \frac{S_0}{K} + \mu \, s}{\sigma_0 \, \sqrt{T}} - \frac{1%
}{2} \sigma_0 \, \sqrt{T} \right] e^{ -\frac{1}{2} \left\{\frac{\ln \frac{S_0%
}{K}+ \mu \, s}{ \sigma_0 \, \sqrt{T}} - \frac{1}{2} \sigma_0 \, \sqrt{T}
\right\}^2} \, ds \\
& = & - K \; \frac{ \mu \, T}{\sqrt{2\, \pi} \, \sigma_0} \int_0^1 \, (1-u)
\; \left[ \frac{ \mu \, T}{\sigma_0 \, \sqrt{T}} u + d_2 \right] exp \left\{
-\frac{1}{2} \left[ \frac{ \mu \, T}{ \sigma_0 \, \sqrt{T}} u + d_2 \right]%
^2\right\} \, ds
\end{eqnarray*}

We integrate by part and we find

\begin{equation}
\mathbb{E} \left[ \int_0^T \frac{\partial^2 F}{\partial \sigma \partial x}%
(\sigma_0, S_s, s) dS_s \right] = - \frac{K \, \sqrt{T}}{\sqrt{2 \, \pi}}
e^{-\frac{1}{2}d_2^2} + \frac{K \, \sqrt{T}}{ \mathcal{L}} \left[\mathcal{N}%
\left( d_2 + \mathcal{L} \right) - \mathcal{N}\left( d_2 \right) \right]
\end{equation}
where $\mathcal{N}$ is the cumulated distribution function of a reduced
gaussian random variable. 

The second term that we have to compute in equation (\ref{eqn:second-corr})
is

\begin{equation}
\mathbb{E} \left[ \int_0^T \frac{\partial^3 F}{\partial \sigma^2 \, \partial
x}(\sigma_0,\, S_s,\, s)\; dS_s \right]
\end{equation}

We can compute this term following the same steps we used for the first term
(\ref{eqn:first-term}).

\begin{eqnarray*}
\mathbb{E} \left[ \int_0^T \frac{\partial^3 F}{\partial \sigma^2 \, \partial
x}(\sigma_0,\, S_s, \, s) dS_s \right] & = & \frac{ \mu}{\sqrt{2 \, \pi}\;
\sigma_0^2} \int_0^T \mathbb{E}\left[ e^{-\frac{1}{2} \, d_1^2(S_s, \, s)}
\; \left\{d_1(S_s, \, s) + d_2(S_s, \, s) - d_1(S_s, \, s)\, d_2^2(S_s,\, s)
\right\} \; S_s \right] ds \\
& = & \frac{ \mu \, S_0}{\sqrt{2\, \pi}\; \sigma_0^2} \int_0^T \mathbb{E}%
\left[ e^{-\frac{1}{2} \, \left[\frac{B_s}{\sqrt{T-s}} + \Theta +\Lambda %
\right]^2 } \, e^{\mu\,s -\frac{1}{2}\sigma_0^2 s + \sigma_0 B_s} \right. \\
& & \star \;\left. \left\{ 2 \left[\frac{B_s}{\sqrt{T_s}} + \Theta - \Lambda%
\right] + 2 \Lambda - \left[\frac{B_s}{\sqrt{T_s}} + \Theta - \Lambda\right]%
^3 -2 \Lambda \left[\frac{B_s}{ \sqrt{T_s}} + \Theta - \Lambda\right]^2
\right\} \right] ds \\
& = & \frac{ \mu \, S_0}{\sqrt{2 \pi} \sigma_0^2} \int_0^T \int_{\mathbb{R}}
e^{-\frac{1}{2} \, \left[\sqrt{\frac{s}{T-s}} y + \Theta +\Lambda \right]^2
} \, e^{\mu\,s -\frac{1}{2}\sigma_0^2 s + \sigma_0 \sqrt{s} y} \frac{e^{-%
\frac{1}{2}y^2}}{\sqrt{2 \, \pi}} \\
& & \star \; \left\{ 2 \left[ \sqrt{\frac{s}{T-s}} y + \Theta - \Lambda%
\right] + 2 \Lambda - \left[\sqrt{\frac{s}{T-s}} y + \Theta - \Lambda\right]%
^3 \right. \\
& & \left. -2 \Lambda \left[\sqrt{\frac{s}{T-s}} y + \Theta - \Lambda\right]%
^2 \right\} \,dy \, ds \\
& = & -\frac{ \mu \, T\, K}{\sqrt{2 \pi} \sigma_0^2} \, \int_0^1 e^{ -\frac{1%
}{2} \left( d_2 + \frac{\mu \, \sqrt{T}}{\sigma_0} \; x \right)^2 } \left\{
2 (1-x) \left(d_2 + \frac{\mu \, \sqrt{T}}{\sigma_0 } \; x \right) \right. \\
& & + \left. \sigma_0 \, \sqrt{T} \left(1-x\right)^2 \left(d_2 + \frac{\mu
\, \sqrt{T}}{\sigma_0 } \; x \right)^2 + \left(1-x\right)^2 \left(d_2 + 
\frac{\mu \, \sqrt{T}}{\sigma_0} \; x\right)^3 \right\} \, dx \\
& = & -\frac{ K\, \sqrt{T} }{\sqrt{2 \pi} \sigma_0} \, \int_{d_2}^{d_2+ 
\frac{\mu \sqrt{T}}{\sigma_0} } e^{ -\frac{1}{2} y^2 } \left\{ 2 \left[1- 
\frac{\sigma_0}{\mu \, \sqrt{T}} (y-d_2)\right] \; y \right. \\
& & + \left. \sigma_0 \, \sqrt{T} \left[1- \frac{\sigma_0}{\mu \, \sqrt{T}}
(y-d_2)\right]^2 \;y^2 + \left[1- \frac{\sigma_0}{\mu \, \sqrt{T}} (y-d_2)%
\right]^2 \; y^3 \right\} \, dy \\
\end{eqnarray*}
where 
\begin{eqnarray*}
\Theta & = & \frac{\ln \frac{S_0}{K} + \mu\, s - \frac{1}{2} \sigma_0^2 s}{%
\sigma_0 \sqrt{T-s}} \\
\Lambda & = & \frac{1}{2} \sigma_0 \sqrt{T-s} \\
\sqrt{\frac{T-s}{T}}\left(\Theta- \Lambda \right) &=& \frac{\ln \frac{S_0}{K}
+ \mu\, s - \frac{1}{2} \sigma_0^2 T}{\sigma_0 \sqrt{T}}= d_2 + \frac{\mu}{%
\sigma_0 \, \sqrt{T}} \; s
\end{eqnarray*}

Finally, we integrate by parts three times and we find

\begin{equation}
\begin{array}{rcl}
\displaystyle \mathbb{E} \left[ \int_0^T \frac{\partial^3 F}{\partial
\sigma^2 \partial x}(\sigma_0, S_s, s) dS_s \right] & = & \displaystyle - 
\frac{K \, \sqrt{T}}{\sigma_0 \,\sqrt{2 \, \pi}} \left\{ \frac{\sigma_0 \, 
\sqrt{T}}{\mathcal{L}} \, \left(1 + \frac{d_2}{ \mathcal{L}}\right) - \frac{8%
}{\mathcal{L}^2}\right\} \, e^{-\frac{1}{2} \left( d_2 + \mathcal{L}
\right)^2} \\ 
\scriptscriptstyle & \scriptscriptstyle & \scriptscriptstyle \\ 
&  & \displaystyle - \frac{K \, \sqrt{T}}{\sigma_0 \,\sqrt{2 \, \pi}}
\left\{ d_2^2 +4 -2 \, \frac{d_2}{\mathcal{L}} + \frac{8}{\mathcal{L}^2 } +
\sigma_0 \, \sqrt{T} \left[ d_2 - \frac{ 4}{\mathcal{L} } - \frac{ d_2}{%
\mathcal{L}^2} \right] \right\} e^{-\frac{1}{2}\, d_2^2} \\ 
\scriptscriptstyle & \scriptscriptstyle & \scriptscriptstyle \\ 
&  & \displaystyle - \frac{K \, \sqrt{T}}{\sigma_0} \left\{ \sigma_0 \, 
\sqrt{T} \, \left[ \frac{3}{\mathcal{L}^2} + \left(1+ \frac{d_2}{ \mathcal{L}
}\right)^2 \right] - \, \frac{8}{\mathcal{L} } - 6 \frac{d_2}{\mathcal{L}^2 }
\right\} \left[\mathcal{N} \left(d_2 + \mathcal{L} \right) - \mathcal{N}
\left(d_2 \right) \right]%
\end{array}%
\end{equation}

\end{document}